\documentclass[aps,prl,twoside,reprint,showkeys,nofootinbib]{revtex4-2}

\usepackage[utf8]{inputenc}
\usepackage{geometry}
\usepackage{graphicx}
\usepackage[dvipsnames]{xcolor}
\usepackage[pdfstartview=XYZ,
bookmarks=true,
colorlinks=true,
linkcolor=blue,
urlcolor=blue,
citecolor=blue,
pdftex,
bookmarks=true,
linktocpage=true, % makes the page number as hyperlink in table of content
hyperindex=true
]{hyperref}
\usepackage{amsmath}
\usepackage{amssymb}
\usepackage{amsthm}
\usepackage{comment}
\usepackage{multirow}
\usepackage[per-mode=symbol]{siunitx} 
\usepackage{lineno}
\usepackage[normalem]{ulem}
\usepackage{cancel}
\newcommand{\orcid}[1]{\href{https://orcid.org/#1}{\includegraphics[width=8pt]{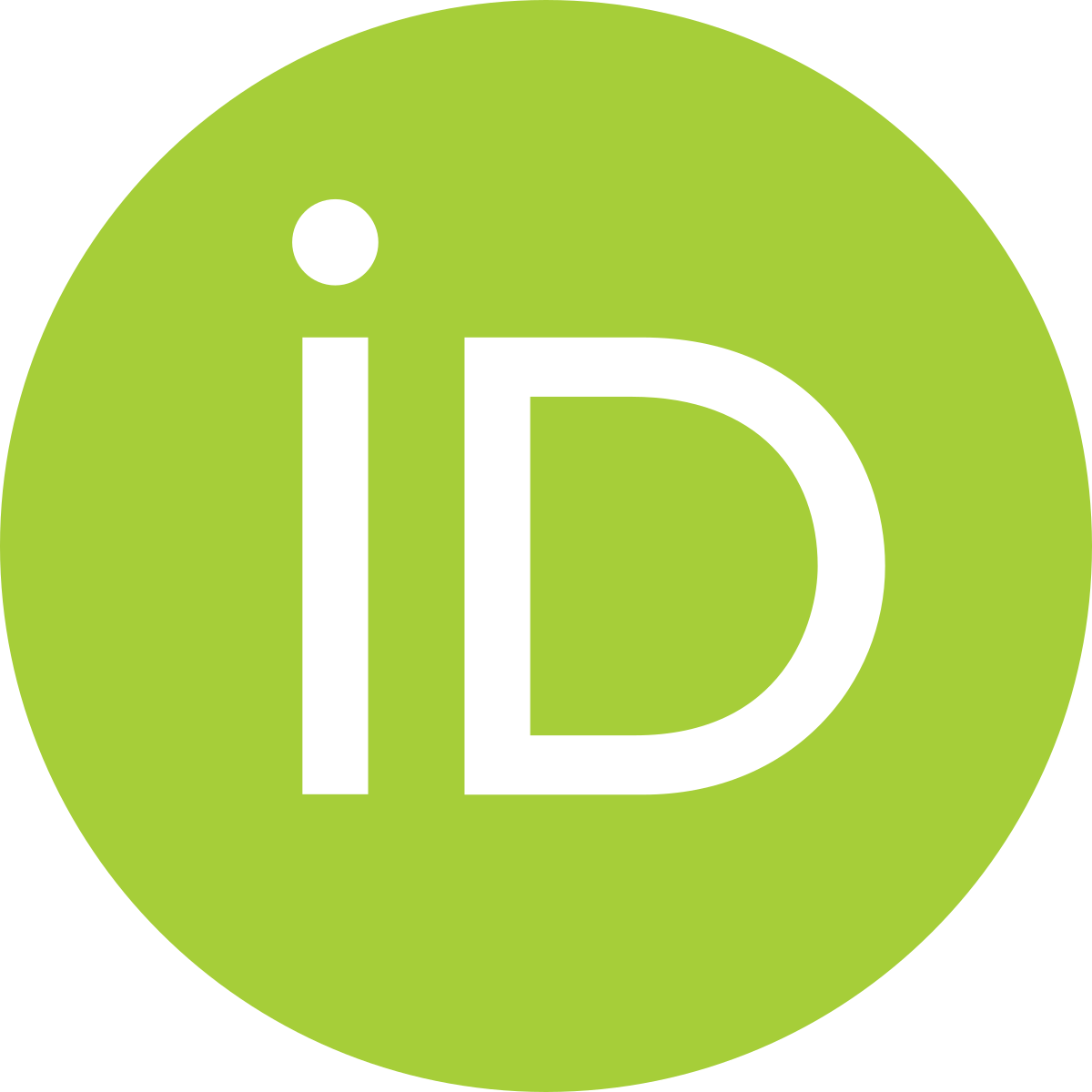}}}
\begin{document}

\title{Updating $\nu_{3}$ lifetime from solar antineutrino spectra}

\author{R. Picoreti\orcid{0000-0001-8306-3098}}

\email{renanpicoreti@gmail.com}
\author{D. Pramanik\orcid{0000-0001-6589-3063}}
\email{dipyaman@ifi.unicamp.br}

\author{P. C. de Holanda\orcid{0000-0001-9852-8900}}
\email{holanda@ifi.unicamp.br}

\author{O. L. G. Peres\orcid{0000-0003-2104-8460}}
\email{orlando@ifi.unicamp.br}
\affiliation{
Instituto de Física Gleb Wataghin - UNICAMP, 13083-859, Campinas SP, Brazil}

\begin{abstract}
We study the production of antineutrinos from the solar neutrinos due the Majorana neutrino decays of neutrino to antineutrino.  Using the antineutrino spectra from KamLAND and Borexino, we present newest limits on the lifetime of $\nu_{3}$ in this scenario. We consider $\nu_{3} \rightarrow \bar{\nu}_{1} + X$ and $\nu_{3} \rightarrow \bar{\nu}_{2} + X$ channels assuming scalar or pseudo-scalar interactions. For hierarchical mass-splittings, the limits obtained by us are $\tau_{3}/m_{3}~\geq~\SI{7d-5}{\second\per\electronvolt}$ and $\tau_{3}/m_{3}~\geq~\SI{1 d-5}{\second\per\electronvolt}$ for the two channels at $90\%$ C.L. We found that the newest bound is five orders of magnitude better than the atmospheric and long-baseline bounds.
\end{abstract}


\maketitle

\section{Introduction}

The question of the neutrino nature, if Dirac or Majorana, is unknown even after several experimental searches. The most known test of neutrino nature is with experiments that search for neutrinoless double-beta decay~\cite{Schechter:1981bd}, which would provide a clear signature  that neutrinos are Majorana particles. However, these experiments did not find evidence for this process and cannot establish the neutrino nature. Other possibilities, for instance, through the search in collision experiments for double muon of same charge signature~\cite{deGouvea:2002gf}, also cannot find any clear signature of the Majorana character of the neutrino.

The transition between flavors during neutrino evolution are detected in different experiments, and provide strong evidence for neutrino masses to be a culprit of this flavor change. Pontecorvo's original idea~\cite{Pontecorvo:1967fh} proposed neutrino to antineutrino transitions, and later it was reformulated to conversion between neutrinos. Therefore, our question is: can we have a different way to produce antineutrinos from a neutrino source that,  if observable, could provide the first signal of Majorana nature? 

\begin{figure}[t]
    \centering
    \includegraphics[width=1.0\columnwidth]{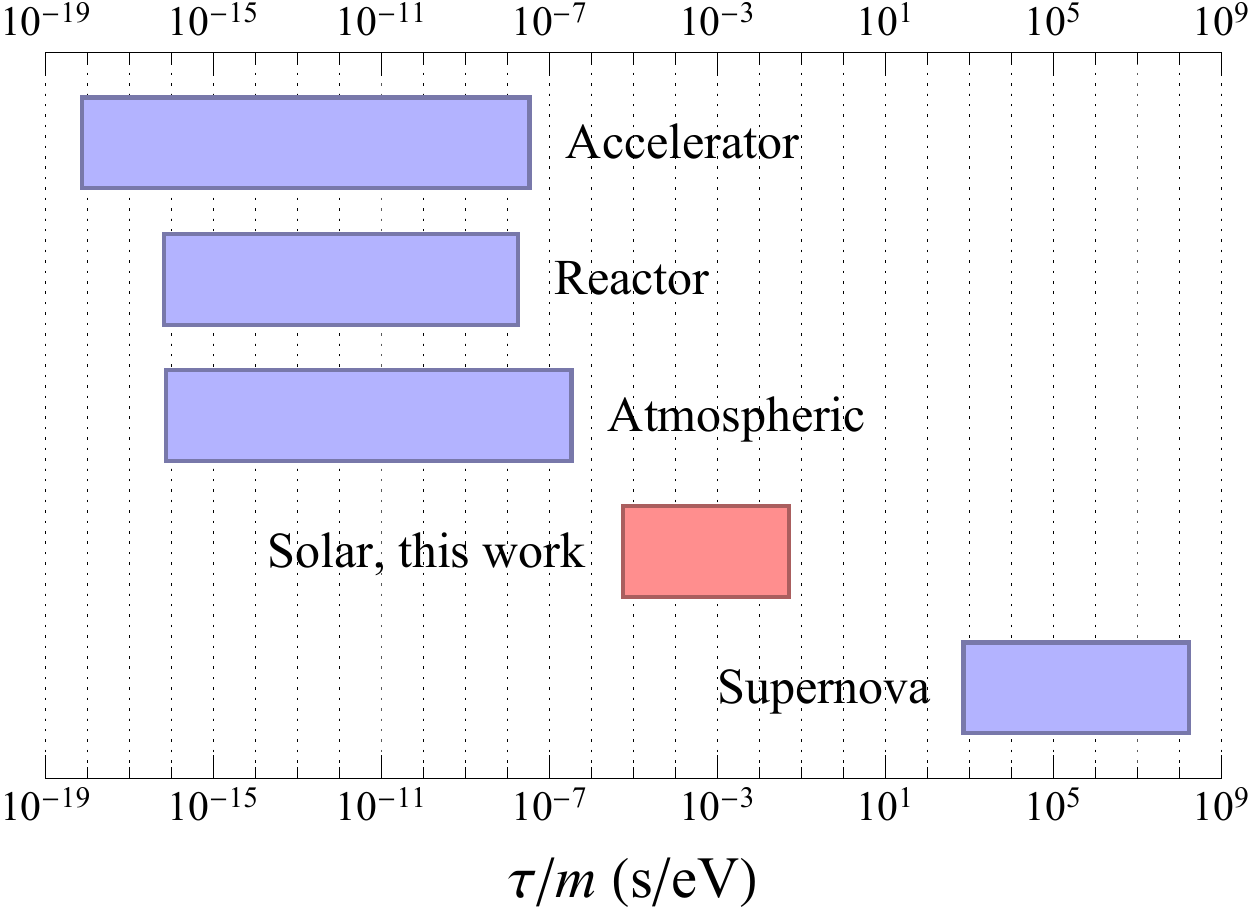}
    \caption{Schematic diagram for the sensitivity of various neutrino sources to the lifetime $\tau/m$ between 99\% and 1\% neutrino survival. Typical baseline values for each source are used.}
    \label{fig:schema}
\end{figure}

Recently there is a lot of discussion about the idea of neutrino decay.
The larger the baseline, or, in other words, the longer the propagation time available for decay, the more sensitive to neutrino decay is the experiment, as shown in Fig.~\ref{fig:schema}. 
 In this scenario a beam of muon neutrinos, that are the linear combination of mass eigenstates,   
$\nu_{\mu}=U_{\mu1}\nu_1+U_{\mu 2}\nu_2+U_{\mu3}\nu_3$ can have its $\nu_3$ component decay into $\nu_{1}$ states for normal ordering. These states are  $\nu_{1}=U_{1e}\nu_e+U_{1\mu}\nu_2+U_{1\tau}
\nu_{\tau}$, and then you can have the appearing of $\nu_{e}$ in the final states. The search for such effect was negative in long-baseline neutrino experiments, atmospheric experiments ~\cite{LoSecco:1998cd,Barger:1998xk,Lipari:1999vh,Fogli:1999qt,Choubey:1999ir,Barger:1999bg,Super-Kamiokande:2004orf,Gonzalez-Garcia:2008mgl,Pakvasa:2012db,Gomes:2014yua,Choubey:2017dyu,Choubey:2017eyg,Choubey:2018cfz,deSalas:2018kri,Tang:2018rer,Ghoshal:2020hyo,Mohan:2020tbi,Chakraborty:2020cfu,Choubey:2020dhw,Hostert:2020oui} where the initial state is richer in  $\nu_{\mu}$ state and it was found the lower bound on neutrino lifetime as $\tau_{3}/m_{3} \geq \SI{2.9d-10}{\second\per\electronvolt}$ at 90\% C.L. Other bounds are possible for reactor neutrinos ~\cite{Abrahao:2015rba,Porto-Silva:2020gma} where it was found that $\tau_{3}/m_{3} \geq \SI{100d-12}{\second\per\electronvolt}$ at 90\% C.L. as for in Ref.~\cite{Porto-Silva:2020gma}. We have given a  summary of these limits in the Supplemental Material.

If neutrinos are Dirac particles, the decay can happen between (anti)neutrino to (anti)neutrino .
If neutrinos are Majorana particles, it can happen through two additional channels, neutrino to antineutrino transition or vice-versa. In the sun, the emission are made of electron neutrinos $\nu_{e}=U_{e 1}\nu_1+U_{e 2}\nu_2+U_{e 3}\nu_3$. If neutrino is a Dirac particle then  from $\nu_3\to \nu_1$ decay you expect a extra $\nu_e$ content, that it is proportional to $U_{e3}
\ll 1$. Otherwise, if the neutrino is a Majorana particle, we can have $\overline{\nu}_1$ component that produces $\overline{\nu}_e$ from the Sun. Then if we search for antineutrinos from the Sun and find a positive result, we can determine the neutrino nature. In this case, the smallness of the $U_{e3}$ is compensated by the large antielectron-neutrino cross-section.
We show that as the standard oscillation produces no antineutrinos, we get robust bounds on $\tau_{3}/m_{3}$ from the solar antineutrino data. This article is organised in the following way. In the first, we describe our model of decay. Next, we show how much antineutrino flux we can get from the Sun for a decaying scenario, and then we present limits from our analysis. Finally, we give our conclusions.

%%%%%%%%%%%%%%%%%%%%%%%%%%%%%%%%%%%%%%%%%%%%%%%%%%%%%%%%%%%%%%%%%%%%%%%%%%%%%
\section{Neutrino Decay Model}

Neutrino masses may arise from the coupling to a scalar singlet known as Majoron~\cite{Chikashige:1980ui, Gelmini:1980re, Coloma:2017zpg}. As a consequence, it is possible for a neutrino to decay into a lighter neutrino alongside the emission of a Majoron. For an interaction Lagrangian with Yukawa scalar and pseudo-scalar couplings, this process is described by~\cite{Lindner:2001fx}
\begin{equation}
\mathcal{L}_\text{int}~=~{\sum_{i,\,j, i\neq j}} (g_s)_{ij}\bar{\nu}_j \nu_i X + i (g_p)_{ij}\bar{\nu}_j\gamma_5 \nu_i X + \text{h.c.}\,,
\end{equation}
where $i,\,j$ are respectively mother and daughter mass eigenstates, while $(g_s)_{ij}$ and $(g_p)_{ij}$ are respectively the scalar and pseudo-scalar coupling constants.

In this work, neutrinos are assumed to be Majorana particles. Neutrinos and antineutrinos are identical and can only be distinguished by their left and right-handed helicities, respectively. Weak interactions couple chiral left-handed neutrinos and chiral right-handed antineutrinos, which, for relativistic neutrinos, are approximated as equal to left, and right helicity states up to terms of order $m/E$. Hence, both left-handed and right-handed Majorana neutrinos are detectable.

The decay rate $\Gamma_{ij}^{rs}$ for each decay process is obtained from the appropriate Feynman diagrams and describe helicity-conserving ($\nu_i^r \rightarrow \nu_j^r$) and helicity-violating ($\nu_i^r \rightarrow \nu_j^s$) decays, where $r,\,s$ denote helicity states. In the following analysis, we assume at each case that only a single heavier active mass eigenstate $\nu_i$ is unstable and decays into neutrinos and antineutrinos of a single lighter active mass eigenstate, $\nu_j$ and $\bar{\nu}_j$. As such 
the energy distribution of the daughter neutrinos as a function of mother and daughter neutrino energies and masses is given by
\begin{equation}
  w_{ij}^{rs}\left(E_i,E_j\right)~=~\frac{1}{\Gamma_{i}^{r}}\frac{d\Gamma_{ij}^{rs} }{dE_j}\left(E_i,E_j\right)\,,
 \end{equation}
such that
\begin{equation}
 w_{ij}^{rs}~=~\left\{
 \begin{array}{ll}
\displaystyle\frac{1}{E_i}\frac{1-A^\pm}{1 - \delta^2}\,, \quad\text{helicity conserving
}\,\\\\
\displaystyle\frac{1}{E_i}\frac{\hspace{0.3cm}A^\pm\hspace{0.3cm}}{1 - \delta^2}\,, \quad\text{helicity violating}\,
 \end{array}\right.\label{eq-dec-distro}
\end{equation}
with the kinematics condition $E_i\,\delta^2\,\leq\,E_j\,<\,E_i$, where $\delta = m_j/m_i$ is the ratio between daughter and mother neutrino masses, in general, $0 \leq \delta < 1$. The function $A^\pm=A^\pm\left(E_i,E_j\right)$ is given by
\begin{equation}\label{eq:A}
 A^\pm~=~\dfrac{1}{\left(1 \pm  \delta\right)^2}\left(1+\delta^2 - \frac{E_j}{E_i}-\delta^2\frac{E_i}{E_j}\right)\,,
\end{equation}
where the plus (minus) sign denotes a scalar (pseudo-scalar) interaction, with ${g_{s}\,\neq\,0}$ and ${g_{p}\,=\,0}$ (with ${g_{s}\,=\,0}$ and ${g_{p}\,\neq\,0}$). We have given a full derivation of the probability in the Supplemental Material.

\begin{figure}[b]
	\centering
	\includegraphics[trim=2cm .5cm 2cm .5cm, clip=true, width=0.9\columnwidth]{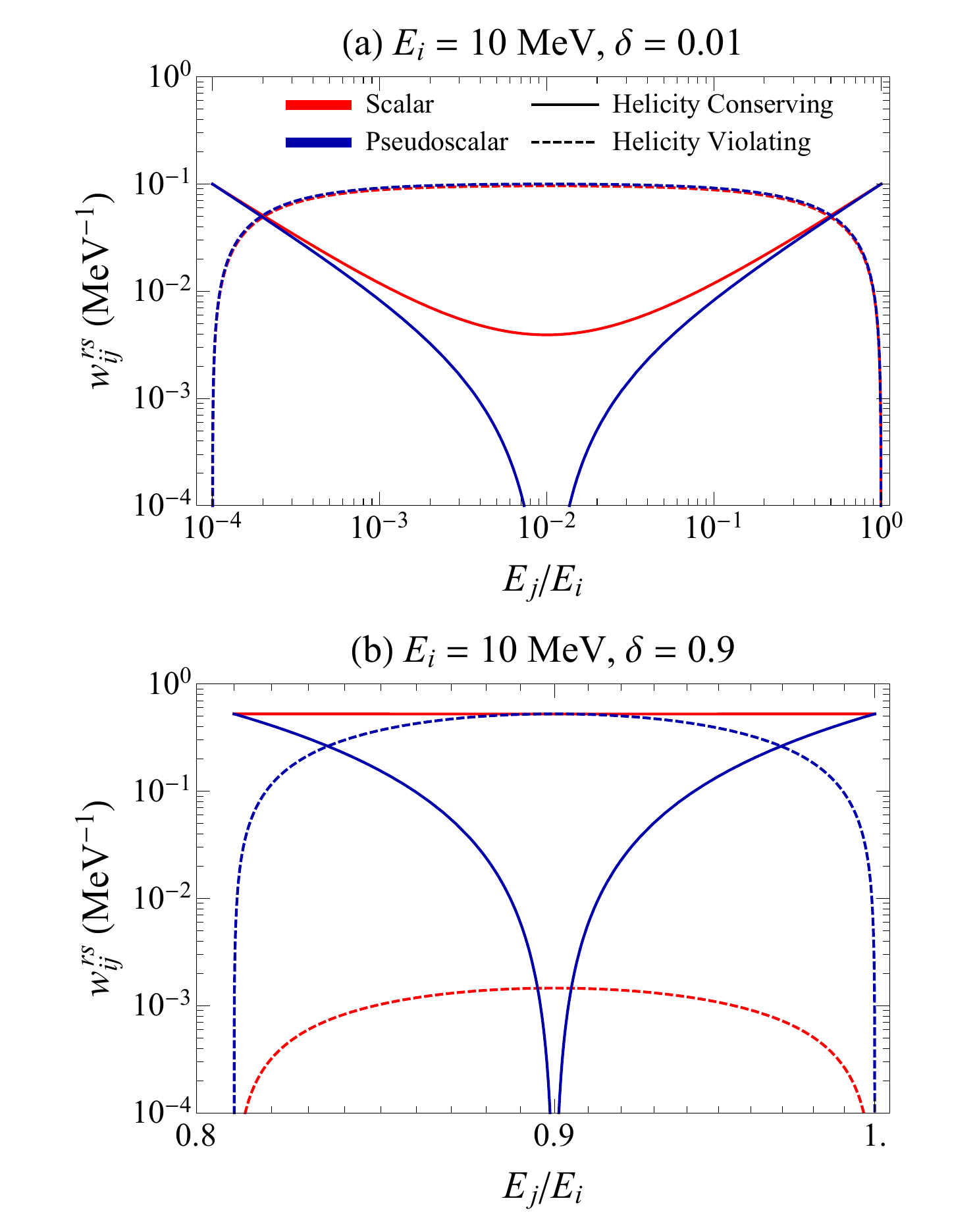}
	\caption{Energy distribution $w_{ij}^{rs}$ of the daughter neutrino or antineutrino $\nu^s_j$ produced in the decay of a $\SI{10}{\mega\electronvolt}$ mother neutrino or antineutrino $\nu^r_i$ for both helicity conserving (solid line) and violating (dashed line) decays as functions of the ratio between daughter and mother neutrino energies $E_j/E_i$ as defined in Eq.~\eqref{eq-dec-distro} for a scalar (red) and pseudoscalar (blue) interaction.}
	\label{fig-dec-distro}
\end{figure}

In  Fig.~\ref{fig-dec-distro}, as $\delta \rightarrow 0$, that is, if neutrino masses are hierarchical, the decays become independent of the coupling constants, and both scalar and pseudoscalar will produce comparable daughter fluxes. On the other hand, as $\delta \rightarrow 1$, if the neutrino masses are quasi-degenerate, the helicity-violating decays are suppressed for the scalar interaction. At the same time, that is not the case for the pseudoscalar interaction where helicity-conserving and violating decays will produce comparable daughter fluxes.

\begin{figure}[t]
    \includegraphics[width=0.95\columnwidth]{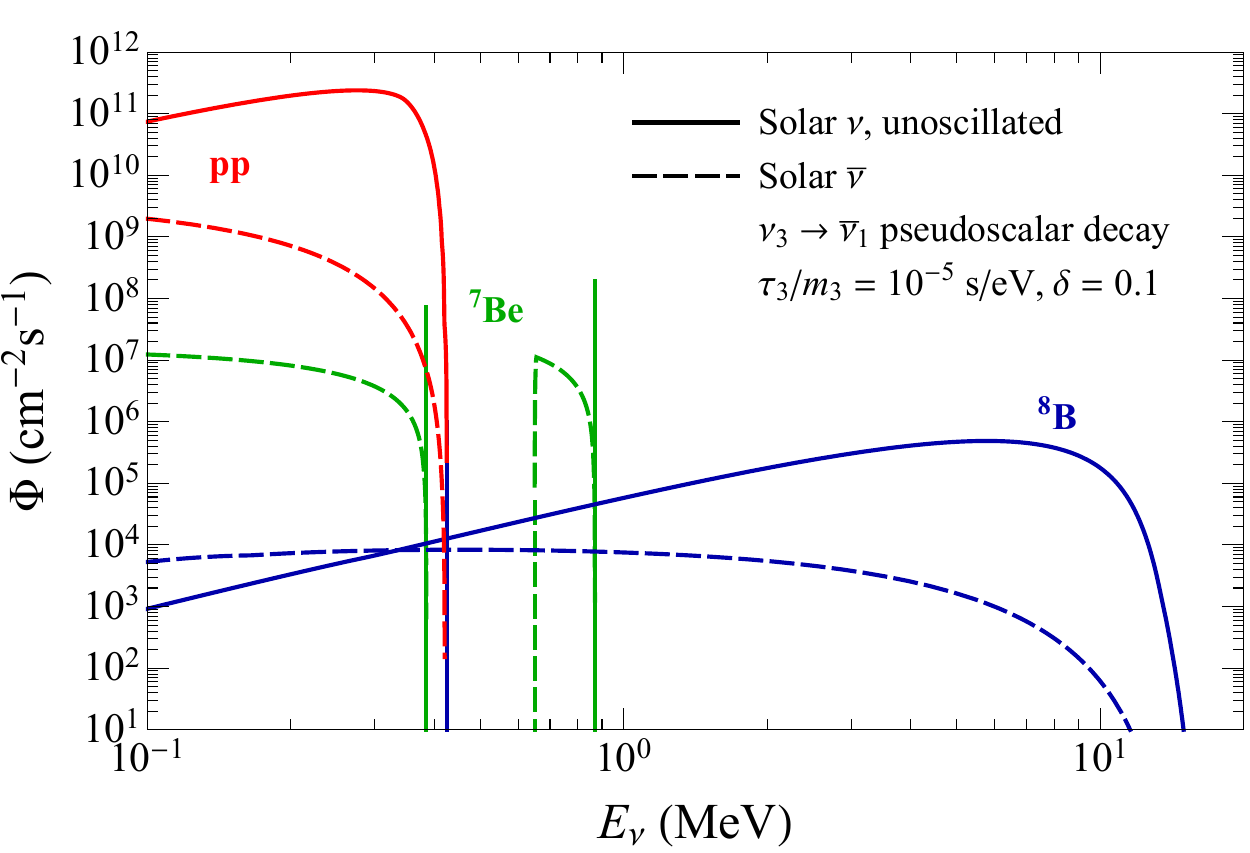}
    \caption{Solar antineutrino $\bar{\nu}_{e}$ flux at Earth due to decay for pseudo-scalar interactions for a $\tau_{3}/m_{3}~=~\SI{e-5}{\second\per\electronvolt}$ assuming $\nu_{3}\rightarrow\bar{\nu}_{1}$ decay. The red, green and the blue curves represent pp, $^7$Be and $^8$B neutrinos respectively. The solid and dashed curves are for the original unoscillated flux, and $\delta=0.1$ respectively.
    }
    \label{fig:flux}
\end{figure}

\section{Antineutrino Flux from decay}
A model-independent combined formalism for obtaining survival and transition probabilities, including neutrino oscillations and decay, is presented in~\cite{Lindner:2001fx}.
Current limits on their lifetime imply that solar neutrinos do not substantially decay either inside the Sun or Earth. As such, assuming solar neutrinos decay only in vacuum on their way from Sun to Earth, the neutrino and antineutrino fluxes arriving at the detector are given by
\begin{widetext}
\begin{equation}
\phi_\beta^{s}(E_j) = \phi_\alpha^r(E_j) \delta_{rs} \displaystyle\sum_{k} P^{\odot}_{\alpha k} \left[\exp\left(-\frac{m_k}{\tau_k}\frac{L}{E_{k}}\right)\right]P^{\oplus}_{k \beta}\, + \displaystyle\int{dE_i\,\phi_\alpha^r(E_i) P^{\odot}_{\alpha i} 
\left[1-\exp\left(-\frac{m_i}{\tau_i}\frac{L}{E_{i}}\right)\right] w^{rs}_{ij}(\delta) P^{\oplus s}_{j \beta} } \label{decay_flux}
\end{equation}
\end{widetext}
with the integration limits $E_j \leq E_i < E_j/\delta^2 $, where $P^{\odot}_{\alpha i}$ is the probability of the produced $\nu_e$ be found as a $\nu_i$ at the surface of the Sun, $P^{\oplus}_{i\beta}$ is the probability of a $\nu_i$ be detected as a $\nu_\beta$ on Earth.
As such, in Eq.~\eqref{decay_flux}, the first term describes the oscillation and decay of the parent neutrinos, while the second term describes the production of daughter neutrinos from the decay of parent neutrinos.

We show the expected $\bar{\nu}_{e}$ flux from the decay channel $\nu_{3}\rightarrow\bar{\nu}_{1}$ for $\tau_{3}/m_{3}\,=\,\SI{e-5}{\second\per\electronvolt}$ as a benchmark value in Fig.~\ref{fig:flux}. The first thing to notice here is that the decay distorts the shape of the flux and pushes the energy of the daughter neutrino towards lower values. As a result, the $^7$Be lines become wider. The broadening of the mono-energetic lines happens because, in two-body decays, the parent's energy is carried by both daughter particles. The kinematic factors determine the width of the line. The daughter energy $E_{j}$ satisfies the conditions $E_{j}\leq E_{i}$ and $E_{j}\,\geq\,E_{i}\delta^{2}$ with $E_{i}$ being the parent neutrino energy. We also see that, for the hierarchical scenario, at ultra-low energies, the expected antineutrino flux can even be larger than the unoscillated flux at those energies, as seen in the $^8$B flux.

\section{Limits from the antineutrino data}
\begin{figure}
    \centering
    \includegraphics[trim=1.2cm .1cm 1.2cm .1cm, clip=true, width=0.95\columnwidth]{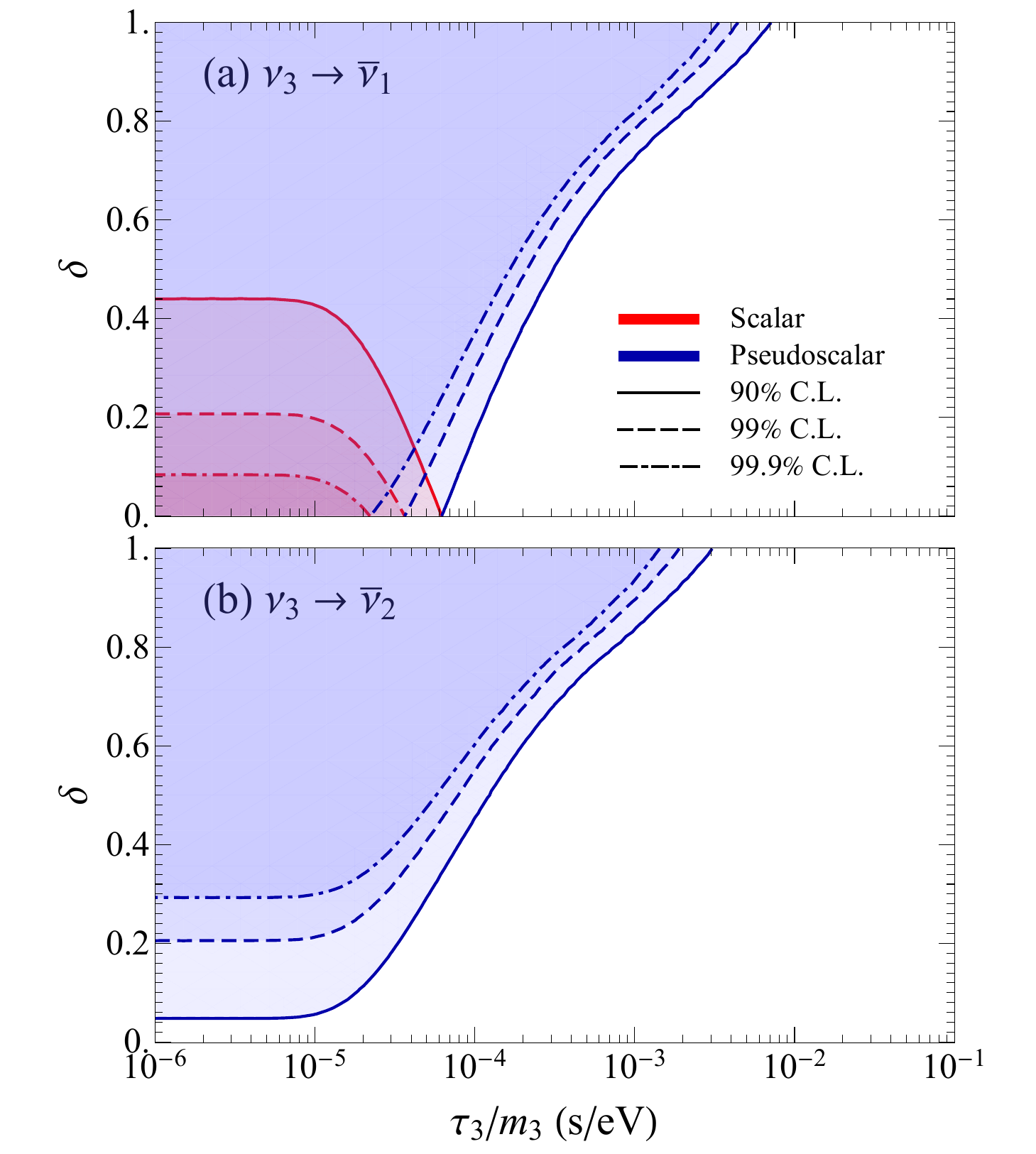}
    \caption{Limits on the $\delta$ vs $\tau_{3}/m_{3}$ plane from the antineutrino data. The shaded regions are disallowed. The red and blue are for scalar and pseudo-scalar interactions respectively with solid, dashed and dashed-dotted lines for $90\%$, $99\%$ and $99.9\%$ confidence levels respectively. The top (bottom) panels are for $\nu_{3}\rightarrow\bar{\nu}_{1}$ ($\nu_{3}\rightarrow\bar{\nu}_{2}$) decay channel.}
    \label{fig:compare_95}
\end{figure}

%%%%%%%%%%%%%%%%%%%%%%%%%%%%%%%%%%%%%%%%%%%%%%%%%%%%%%%%%%%%%%       Statistical results
%%%%%%%%%%%%%%%%%%%%%%%%%%%%%%%%%%%%%%%%%%%%%%%%%%%%%%%%%%%%
We present limits on the neutrino decay from the solar antineutrino data in Fig.~\ref{fig:compare_95} for the two channels for $\nu_{3}$ decay to antineutrinos, $\bar{\nu}_{1}$ or $\bar{\nu}_{2}$ for the antineutrino data from KamLAND~\cite{Collaboration:2011jza} \footnote{When we were finalizing the draft KamLAND collaboration has published a new result~\cite{Abe:2021tkw}.} and Borexino~\cite{Agostini:2019yuq}. We present here only the limits for the decay of $\nu_{3}$, because getting limits on $\nu_{3}$ lifetime from the solar experiments is a completely novel idea. For other possible channels we have shown the results in the supplementary material. Both KamLAND and Borexino use inverse-beta decay to detect the antineutrinos, and thus are limited by its threshold. Therefore, we only use the $^8$B neutrinos for our analysis as the hep neutrino flux is much smaller than the $^8$B.

To simulate the antineutrino spectra of KamLAND, we have matched the 90\% upper limit of the total number of events given in Ref.~\cite{Collaboration:2011jza} assuming the model of antineutrino conversion probability as given in Ref.~\cite{Collaboration:2011jza}. We assumed a fiducial mass of 1 kt. The data corresponds to 23445 days of exposure. We also ignored the systematic uncertainties and the effect of the finite resolution. To simulate Borexino, like KamLAND analysis we again matched their 90\% C.L. results from Ref.~\cite{Agostini:2019yuq} which corresponds to 2485 days of exposure to a total $1.32\times10^{31}$ number of target nuclei. Again we have ignored the systematic uncertainties and the effect of the finite resolution. To obtain the limits, we have assumed that $\theta_{12}$, $\theta_{13}$, and $\Delta m_{21}^2$ are unknown. So, we have varied these parameters with prior terms according to~\cite{Esteban:2020cvm}. We see in Fig.~\ref{fig:compare_95} that the behaviour of the scalar and the pseudoscalar cases are different. The scalar hypothesis is ruled out only for the $\nu_{3}\rightarrow\bar{\nu}_{1}$ channel for the hierarchical neutrino masses and very fast decay, but for the pseudoscalar interactions, the limit is increased with $\delta$, .i.e, with reducing the mass-splitting between the parent and daughter states. Another interesting thing to note is that both scalar and pseudoscalar interactions give similar bounds for the hierarchical limit.
In the $\delta\rightarrow0$ limit,  the decay-rates becomes independent of the nature of the interactions as can be seen  in  Eq.~\eqref{eq-dec-distro} and Eq.~\eqref{eq:A}. A bound on $\tau_3/m_3$ is estimated in Reference~\cite{Funcke:2019grs} for degenerate (hierarquical) case based on scaling a limit obtained for $\tau_2/m_2$ to account for a small $U_{e3}$ to be $1.3 \times  10^{-4}$ ($2.2 \times 10^{-5}$)~s/eV.

We can understand the behaviour of the scalar and pseudoscalar cases from Fig.~\ref{fig-dec-distro}. We notice that the weighted differential rates are more significant for the pseudoscalar interactions than the scalar interactions for the helicity-violating decays, which are responsible for the antineutrino appearances. Thus we find weaker limits for the scalar scenario than for the pseudoscalar scenario. By comparing two panels of Fig.~\ref{fig-dec-distro}, it becomes clear that a higher decay rate for the scalar case happens for lower values of $\delta$.  The decay largely diminishes as the $\delta$ increases, and there are no limits for the quasi-degenerate region in the case of scalar interactions. However, the weighted differential decay rate for the pseudoscalar case increases as we go from the hierarchical to the quasi-degenerate region. As a result, we observe the limits getting stronger as we increase $\delta$ for the pseudoscalar case. 

We also see that $\nu_{3}\rightarrow\bar{\nu}_{1}$ decay gives better bounds than $\nu_{3}\rightarrow\bar{\nu}_{2}$ decay. In fact the limit is so poor for latter case, that we don't see any limit for the scalar case and for hierarchical scenario for the pseudo-scalar case. From the second term of the Eq.~\eqref{decay_flux}, we note that the probability depends on the $P^{\oplus}_{ie}$, where $i$ is the daughter neutrino mass-eigenstate. Now, $P^{\oplus}_{1e}=c^{2}_{13}c^{2}_{12}$ and $P^{\oplus}_{2e}=c^{2}_{13}s^{2}_{12}$, so $\theta_{12}\sim33.5^{\circ}$ makes $P^{\oplus}_{1e}\simeq0.67$ and $P^{\oplus}_{2e}\simeq0.33$. Hence, $\nu_{3}\rightarrow\bar{\nu}_{1}$ gives more antineutrinos compared to $\nu_{3}\rightarrow\bar{\nu}_{2}$ decay and thus stronger limits for the first channel. 

\section{Conclusion}
 In this letter we present an analysis of solar-antineutrino spectra at KamLAND and Borexino through the decay of a heavier neutrino state into a lighter antineutrino and a Majoron. Previously, neutrino data from the sun could only constrain $\nu_{2}$ decay, but searches for the solar antineutrino spectra by experiments like KamLAND and Borexino, together with a positive measurement of $\theta_{13}$, has enabled us to look for $\nu_{3}$ decay. We consider two channels $\nu_{3}\rightarrow\bar{\nu}_{1} + X$ and $\nu_{3}\rightarrow \bar{\nu}_{2}+X$, both with purely scalar interactions and purely pseudo-scalar interactions. We study them as a function of the mass-splitting $\delta$ between the parent and daughter neutrino states. To put our main results in a nutshell we present here limits at $90\%$ C.L. for two benchmark values of $\delta$, $\delta=0.2$ and $\delta=0.8$. For the $\nu_{3}~\rightarrow~\bar{\nu}_{1}~+~X$ decay channel with a pseudo-scalar interaction, we obtain the limits $\tau_{3}/m_{3}~\geq~\SI{e-4}{\second\per\electronvolt}$ and $\tau_{3}/m_{3}~\geq~\SI{2e-3}{\second\per\electronvolt}$ for the two benchmark $\delta$ values, respectively. For scalar antineutrino interaction, data does not put limits on the larger values of $\delta$. For $\delta=0.2$ the limit is $\tau_{3}/m_{3}~\geq~\SI{3e-5}{\second\per\electronvolt}$. For $\nu_{3}~\rightarrow~\bar{\nu}_{2}~+~X$ decay channel, we do not get any limit for the scalar case however the limits for the pseudo-scalar case for the two benchmark $\delta$'s values are $\tau_{3}/m_{3}~\geq~\SI{3e-5}{\second\per\electronvolt}$ and $\tau_{3}/m_{3}~\geq~\SI{e-3}{\second\per\electronvolt}$ respectively.
 
 To conclude, a positive measurement of solar antineutrino spectra would open up a new window of possibilities. As standard neutrino physics predicts no solar antineutrino, any observation of antineutrino from the Sun will be a new physics signal. There can be a plethora of novel ideas that can be tested using the solar antineutrino data which we leave for future work.

\section*{Acknowledgements}
P.C.H., O.L.G.P. and D.P. were thankful for the support of FAPESP funding Grant 2014/19164-6. O.L.G.P were thankful for the support of CNPq grant 306565/2019-6. D.P. is thankful for the support of FAPESP fellowship  2020/04261-7. This study was financed in part by the Coordenação de Aperfeiçoamento de Pessoal de Nível Superior - Brasil (CAPES) - Finance Code 001.

\bibliographystyle{apsrev4-2}
\bibliography{refs}

\onecolumngrid
\appendix
\label{j1}
\ifx \standalonesupplemental\undefined
\setcounter{page}{1}
\setcounter{figure}{0}
\setcounter{table}{0}
\setcounter{equation}{0}
\fi

\renewcommand{\thepage}{Supplemental Material-- S\arabic{page}}
\renewcommand{\figurename}{SUPPL. FIG.}
\renewcommand{\tablename}{SUPPL. TABLE}

\renewcommand{\theequation}{A\arabic{equation}}
%\clearpage

\begin{center}
\textbf{\large Supplemental Material}
\end{center}

\section{Neutrino Decay Model}
\label{supple}
We assume the parent neutrino decays into a lighter active neutrino and a Majoron~\cite{Lindner:2001fx} --- $\nu_{i}\rightarrow\nu_{j}+X$~--- for which the decay width is given by
\begin{equation}
d\Gamma = \frac{1}{2E_i}\left|\mathcal{M}\right|^2(2\pi)^4\delta^{(4)}(\mathbf{p}_i-\mathbf{p}_j-\mathbf{p}_X)
\frac{d^3p_j}{(2\pi)^3}\frac{1}{E_j}\frac{d^3p_{X}}{(2\pi)^3}\frac{1}{E_X}\,,
\end{equation}
where $\mathbf{p}_i = (E_{i}, \vec{p}_{i})$, $\mathbf{p}_j = (E_{j}, \vec{p}_{j})$, $\mathbf{p}_X = (E_{X}, \vec{p}_{X})$ are respectively $\nu_i$, $\nu_j$ and $X$ four-momenta, with ${E_{i}^2 = |\vec{p}_{i}|^2 + m_{i}^2}$,  ${E_{j}^2 = |\vec{p}_{j}|^2 + m_{j}^2}$ and ${E_{X}^2 = |\vec{p}_{X}|^2 + m_{X}^2}$. In this analysis, we suppose a \textit{massless} Majoron, that is, $m_{X} = 0$. The matrix elements $\left|\mathcal{M}\right|^2$ are given by
\begin{align}
   \left|\mathcal{M}^{rr}_{ij}\right|^{2} =& \frac{g_{s}^{2}}{4}(A+2)+\frac{g_{p}^{2}}{4}(A-2)\,,\\
   \left|\mathcal{M}^{rs}_{ij}\right|^{2} =& \frac{g_{s}^{2}+g_{p}^{2}}{4}\bigg(\frac{1}{\delta} + \delta-A\bigg)\,,
\end{align}
where $g_s = (g_s)_{ij}$ and $g_p = (g_p)_{ij}$ are respectively the scalar and pseudoscalar coupling constants, $\delta~=~m_j/m_i$, and $r,s$ denote helicity states, such that $rr$ ($rs$) implies a helicity conserving (violating) interaction, and:
\begin{equation}
    A = \delta\frac{E_{i}}{E_{j}} + \frac{1}{\delta}\frac{E_{j}}{E_{i}}\,.
\end{equation}
The differential decay width is given by
\begin{equation}
    \frac{d\Gamma^{rs}_{ij}}{dE_{j}} = \frac{m_{i}m_{j}}{4\pi E^{2}_{i}}\bigg(1-\frac{m^{2}_{i}}{E^{2}_{i}}\bigg)\left|\mathcal{M}^{rs}_{ij}\right|^{2},
\end{equation}
with the kinematic conditions constraining the energies and the angle between initial and the final neutrinos
\begin{equation}
    E_{i}-E_{j}=\big(|\vec{p}_{i}|^{2}+|\vec{p}_{j}|^{2}-2|\vec{p}_{i}||\vec{p}_{j}|\cos\theta\big)^{1/2}\,.
\end{equation}
For the ultra-relativistic neutrinos, this results in bounds on the energy on the daughter neutrino as
\begin{equation}
    E_{i}\delta^{2}\leq E_{j}\leq E_{i}\,.
\end{equation}
For the interaction above, we obtain the decay widths for the helicity conserving and violating decays respectively as
\begin{align}
    \Gamma^{rr}_{ij} =& \frac{m^{2}_{i}}{32\pi E_{i}}\left[\big(g_{s}^{2}+g_{p}^{2}\big)\big(1-4\delta^{2}\ln\delta-\delta^{4}\big)+(g_{s}^{2}-g_{p}^{2}\big)4\delta\big(1-\delta^{2}\big)\right]\,,\\
    \Gamma^{rs}_{ij} =& \frac{m^{2}_{i}}{32\pi E_{i}}\left[\big(g_{s}^{2}+g_{p}^{2}\big)\big(1+4\delta^{2}\ln\delta-\delta^{4}\big)\right]\,.
    \end{align}
Finally the neutrino lifetime can be written in terms of the decay widths as
\begin{equation}
    \frac{m_{i}}{\tau_{i}} = E_{i}\Gamma^{r}_{i} = E_{i}\sum\limits_{k,s}\Gamma^{rs}_{ik}\,.
\end{equation}
Under the assumption of a single heavier neutrino decaying into a single lighter daughter, the total decay width simplifies to 
\begin{equation}
    \Gamma^{r}_{i} = \frac{m^{2}_{i}}{16\pi E_{i}}\big[\big(g_{s}^{2}+g_{p}^{2}\big)\big(1-\delta^{4}\big)+\big(g_{s}^{2}-g_{p}^{2}\big)2\delta\big(1-\delta^{2}\big)\big].
\end{equation}
Next, by dividing the differential decay widths by the total decay width we obtain the energy distribution of the daughter neutrinos as
\begin{align}
w^{rr}_{ij} =& \frac{1}{\Gamma^{r}_{i}}\frac{d\Gamma^{rr}_{ij}}{dE_{j}} = \frac{f_{1}}{E_{i}} -\frac{f_{2}}{E_{i}}\bigg(f_{1}\mp\frac{m^{2}_{i}g^{2}_{s(p)}}{8\pi E_{i}\Gamma^{r}_{i}}f_{3}\bigg)\bigg(1+\delta^{2}-\frac{E_{j}}{E_{i}}-\delta^{2}\frac{E_{i}}{E_{j}}\bigg), \label{en-distro-1}\\
w^{rs}_{ij} =& \frac{1}{\Gamma^{r}_{i}}\frac{d\Gamma^{rs}_{ij}}{dE_{j}} = \frac{f_{2}}{E_{i}}\bigg(f_{1}\mp\frac{m^{2}_{i}g^{2}_{s(p)}}{8\pi E_{i}\Gamma^{r}_{i}}f_{3}\bigg)\bigg(1+\delta^{2}-\frac{E_{j}}{E_{i}}-\delta^{2}\frac{E_{i}}{E_{j}}\bigg), \label{en-distro-2}
\end{align}
where $f_{1,s(p)} = 1/(1-\delta^{2})$, $f_{2,s(p)} = 1/(1\pm\delta^{2})$ and $f_{3,s(p)} = 2\delta$, and the upper (lower) sign corresponds to $g_{s}(g_{p})$. From Eqs.~\eqref{en-distro-1} and \eqref{en-distro-2} we can obtain the energy distribution for a purely scalar interaction ($g_{p}=0$) or purely pseudo-scalar interaction ($g_{s}=0$) as
\begin{align}
    w^{rr}_{ij} =& \frac{1}{\big(1-\delta^{2}\big)}\frac{1}{E_{i}}-\frac{1}{\big(1-\delta^{2}\big)\big(1\pm\delta\big)^{2}}\frac{1}{E_{i}}\bigg(1+\delta^{2}-\frac{E_{j}}{E_{i}}-\delta^{2}\frac{E_{i}}{E_{j}}\bigg),\\
    w^{rs}_{ij} =& \frac{1}{\big(1-\delta^{2}\big)\big(1\pm\delta\big)^{2}}\frac{1}{E_{i}}\bigg(1+\delta^{2}-\frac{E_{j}}{E_{i}}-\delta^{2}\frac{E_{i}}{E_{j}}\bigg).
\end{align}
where the upper (lower) sign corresponds to a purely scalar (pseudoscalar) interaction. Finally, for the sake of completeness, the branching ratios are given by
\begin{align}
    {\rm Br}^{rr}_{ij} = & \frac{1+\delta^{2}}{2\big(1+\delta\big)^{2}}-\frac{2\delta^{2}\ln\delta}{\big(1\pm\delta\big)^{2}\big(1-\delta^{2}\big)}+\frac{2\delta}{\big(1\pm\delta\big)^{2}},\\
    {\rm Br}^{rs}_{ij} = & \frac{1+\delta^{2}}{2\big(1+\delta\big)^{2}}+\frac{2\delta^{2}\ln\delta}{\big(1\pm\delta\big)^{2}\big(1-\delta^{2}\big)}.
\end{align}
where, again, the upper (lower) sign corresponds to a purely scalar (pseudoscalar) interaction.

\section{Oscillation probability under decay}
Here we follow the formalism developed in Ref.~\cite{Lindner:2001fx} to get the oscillation probability involving decay. We introduce three operators in terms of the creation and annihilation operators. The first one is the propagation operator. This gives the amplitude of propagation of a state of energy $E_{i}$ for a distance $l$. We define this as:
\begin{equation}\label{eq:El}
    \mathcal{E}(l) = \sum\limits_{i}\exp\left(-iE_{i}l\right)\hat{a}^{r\dagger}_{i}\hat{a}^{r}_{i}.
\end{equation}
Where $\hat{a}^{r}_{i}$ destroys a state in mass-eigenstate $i$ with helicity $r$ and $\hat{a}^{r\dagger}_{i}$ creates a state in mass-eigenstate $i$ with helicity $r$. Next we define the disappearance operator which gives the amplitude of a neutrino state remaining undecayed after propagating a distance of $l$ along its baseline. It is defined as:
\begin{equation}\label{eq:Dl}
    \mathcal{D}_{-}(l) = \sum\limits_{i}\exp\left(-\frac{\Gamma^{r}_{i}l}{2}\right)\hat{a}^{r\dagger}_{i}\hat{a}^{r}_{i}.
\end{equation}
Finally the appearance operator describes destruction of state $i$ with energy $E_{i}$ and chirality $r$, and creation of the daughter state in $j$th eigenstate with energy $E_{j}$ with chirality $s$ between a distance of $l$ and $l+dl$ along the baseline.
\begin{equation}\label{eq:D+l}
    \mathcal{D}_{+}(l) = \sum_{i,j,i\neq j}\left(\Gamma^{rs}_{ij}\eta_{ij}\right)^{1/2}\exp\left(i\xi\right)\hat{a}^{s\dagger}_{j}\hat{a}^{r}_{i}.
\end{equation}
Where, $\Gamma^{rs}_{ij}$ gives the decay rate for the transition between $i$th to $j$th state, $\xi$ is a random phase due to the phase shifts caused by other invisible final state particles produced in decay, and $\eta^{rs}_{ij}$ is the fraction of the decay products that pass through the detector given by
\begin{equation}
    \eta^{rs}_{ij}(l,L,D) =
    \frac{1}{\Gamma^{rs}_{ij}}\int^{E_{\rm max}}_{E_{\rm min}}\int^{1}_{\cos\theta_{D}}\left|\frac{d\Gamma^{rs}_{ij}}{d\cos\theta dE_{j}}\big(E_{i},E_{j}\big)\right|dE_{j}d\cos\theta.
\end{equation}

As most of the neutrinos are produced in the core of the Sun and the Sun is radially symmetric, the the neutrino production zone is observed from the Earth within a very small angle. Also, the decay of the ultra-relativistic neutrinos are highly forward peaked. Hence, to a first approximation almost all neutrinos reach the detector in the desired energy range. Thus $\eta^{rs}_{ij}$ can be approximated as 
\begin{equation}
    \frac{d\eta^{rs}_{ij}}{dE_{j}} = \frac{1}{\Gamma^{rs}_{ij}}\frac{d\Gamma^{rs}_{ij}}{dE_{j}}\big(E_{i},E_{j}\big).
\end{equation}

Next, $P^{(n)}_{ij}$ is defined as transition probability between $i$th and $j$th mass-eigenstates with n-intermediate states. For $n=0$, this is given by
\begin{equation}
    P^{(0)}_{ij} = \left|\langle\nu^{s}_{j}|\mathcal{E}(L)\mathcal{D}_{-}(L)|\nu^{r}_{i}\rangle\right|^{2}.
\end{equation}
This is the case, where the initial particle has disappeared and no new particle has appeared. Therefore $P^{0}_{ij}$ gives the transition probability for invisible decays. In terms of flavour state transition from $\nu_{\alpha}$ to $\nu_{\beta}$ along a baseline $l$, the probability is given by
\begin{equation}
    P^{\rm inv}_{\alpha\beta} = \bigg|\sum\limits_{i}\sum\limits_{j}\big(V^{r}_{\alpha i} \big)^{*}\big(V^{r}_{\beta j}\big)\langle\nu^{s}_{j}|\mathcal{E}(L)\mathcal{D}_{-}(L)|\nu^{r}_{i}\rangle\bigg|^{2}\,,
\end{equation}
where $V$ is the usual PMNS matrix. From Eq.~\eqref{eq:El} and Eq.~\eqref{eq:Dl} we obtain
\begin{equation}
    P^{\rm inv}_{\alpha\beta} = \bigg|\sum\limits_{i}\big(V^{r}_{\alpha i} \big)^{*}\big(V^{r}_{\beta i}\big)\exp\bigg(-\frac{\tilde{m}^{2}_{i}l}{2E_{i}}\bigg)\bigg|^{2}\,,
\end{equation}
where $\tilde{m_i}^2 = i m_i^2 + E_i \Gamma_i^r$. 

For $n>0$ the probability is given by
\begin{equation}
    P^{(n)}_{ij} = \int^{L}_{0}dl_{1}...\int^{L}_{l_{n}-1}dl_{n}\bigg|\langle\nu^{s}_{j}|\mathcal{E}(L-l)\mathcal{D}_{-}(L-l)\bigg[\prod_{i=1}^{n}\mathcal{D}_{+}(l_{i})\mathcal{E}(l_{i})\mathcal{D}_{-}(l_{i})\bigg]|\nu^{r}_{i}\rangle\bigg|^{2},
\end{equation}
with $l=\sum\limits_{i=1}^{n}l_{i}$. 

For the visible decay we assume that there is only one intermediate state and neglect $n>1$. Therefore the probability for visible decay is 
\begin{equation}\label{eq:vis0}
    P^{\rm vis}_{ij} = P^{(0)}_{ij}+P^{(1)}_{ij}, 
\end{equation}
with
\begin{equation}
    P^{(1)}_{ij} = \int^{L}_{0}dl\left|\langle\nu^{s}_{j}|\mathcal{E}(L_l)\mathcal{D}_{-}(L-l)\mathcal{D}_{+}(l,L)\mathcal{E}(l)\mathcal{D}_{-}(l)|\nu^{r}_{i}\rangle\right|^{2}.
\end{equation}
Using Eqs.~\eqref{eq:El}, \eqref{eq:Dl} and \eqref{eq:D+l} we get the transition from flavour state $\nu_{\alpha}$ to $\nu_{\beta}$ as
\begin{equation}
    P^{(1)}_{\alpha\beta} = \int^{L}_{0}dl\bigg|\sum\limits_{i,j,i\neq j}\big(V^{r}_{\alpha i}\big)^{*}\big(V^{r}_{\beta j}\big)\left(\Gamma^{rs}_{ij}\eta^{rs}_{ij}\right)^{1/2}\exp\bigg[-\frac{\Tilde{m}^{2}_{i}}{2E_{i}}l-\frac{\Tilde{m}^{2}_{j}}{2E_{j}}(L-l)\bigg]\bigg|^{2}.
\end{equation}
Finally collecting all the pieces together the general expression for the probability under decay becomes 
\begin{align}
    \frac{dP^{\rm rs}_{\alpha\beta}}{dE_{j}} &= \bigg|\sum\limits_{i}\big(V^{r}_{\alpha i }\big)^{*}\big(V^{r}_{\beta i}\big)\exp\bigg(-\frac{\Tilde{m}^{2}_{i}l}{2E_{i}}\bigg)\bigg|^{2}\delta(E_{i}-E_{j})\delta_{rs}\nonumber\\ 
    &+ \int^{L}_{0}dl\bigg|\sum\limits_{i,j,i\neq j}\big(V^{r}_{\alpha i}\big)^{*}\big(V^{r}_{\beta j}\big)\left(\Gamma^{rs}_{ij}\eta^{rs}_{ij}\right)^{1/2}\exp\bigg[-\frac{\Tilde{m}^{2}_{i}}{2E_{i}}l-\frac{\Tilde{m}^{2}_{j}}{2E_{j}}(L-l)\bigg]\bigg|^{2}.
\end{align}
Now, we are in the position to discuss the case of solar neutrino decay. The existing limits on the neutrino lifetime exclude the possibility of large amount of decay inside the Sun or Earth. Hence we only consider the case where neutrinos decay after exiting the Sun and when they are in vacuum. Under these assumption, the evolution of neutrinos inside the Sun follow the standard MSW mechanism. Therefore we can write the first part of the probability as
\begin{equation}
    P^{\rm inv}_{e\beta} = \bigg|\sum\limits_{i}A^{\odot}_{ei}A^{\oplus}_{i\beta}\exp\bigg(-i\frac{m_{i}L}{2E}\bigg)\exp\bigg(-\frac{\Gamma_{i}L}{2}\bigg)\bigg|^{2},
\end{equation}
where $A^{\odot}_{ei}$ denotes the amplitude of a $\nu_{e}$ generated in the Sun to be in state $\nu_{i}$ at the surface. Similarly, $A^{\oplus}_{i\beta}$ represents the transition amplitude of a neutrino eigenstate $\nu_{i}$ to be detected in the Earth as $\nu_{\beta}$ flavour state. Due to the large distance between the Sun and the Earth, the interference terms will be averaged to zero and the probabilities will be the incoherent sum of the square of the amplitudes and this can be written as
\begin{equation}
    P^{inv}_{e\beta} = \sum\limits_{i}P^{\odot}_{ei}P^{}_{ii}P^{\oplus}_{i\beta},
\end{equation}
where $P_{ii}=\exp\left(-\Gamma_{i}L\right)$ for $\nu_{i}$ being the unstable mass-eigenstate and for the rest it is unity.

Next we consider the second term of the visible decay probability. Like the invisible this term can also be written as
\begin{equation}
    \frac{dP^{rs}_{e\beta}}{dE_{j}}=\int^{L}_{0}dl\bigg|\sum\limits_{i,j,i\neq j}A^{\odot}_{ei}A^{\oplus}_{j\beta}\bigg(\frac{d\Gamma^{rs}_{ij}}{dE_{j}}\bigg)^{1/2}\exp\bigg[-\frac{\Tilde{m}^{2}_{i}}{2E_{i}}l-\frac{\Tilde{m}^{2}_{j}}{2E_{j}}(L-l)\bigg]\bigg|^{2}.
\end{equation}
Now if we assume that only $i$th mass eigenstate is unstable, then following similar argument as before we can write the integral as
\begin{equation}
    \frac{dP}{dl} = P^{\odot}_{ei}P^{\oplus}_{j\beta}\frac{d\Gamma^{rs}_{ij}}{dE_{j}}\exp\left(-\Gamma_{i}L\right)\,.
\end{equation}
Integrating we get and substituting we get the probability as
\begin{equation}
    \frac{dP^{rs}_{e\beta}}{dE_{i}} = \sum\limits_{k}P^{\odot}_{ek}P^{}_{kk}P^{\oplus}_{k\beta}\delta(E_{i}-E_{j})\delta_{rs}+P^{\odot}_{ei}P_{ij}w^{rs}_{ij}P^{\oplus s}_{j\beta}\,,
\end{equation}
with $P^{}_{ij} = 1-\exp\left(-\Gamma_{i}L\right)$. Finally, the neutrino flux for the flavour state is given by
\begin{align}
    \phi^{s}_{\beta}(E_{j}) =& \int\phi^{r}_{e}(E_{j})\frac{dP^{rs}_{e\beta}}{E_{i}}dE_{i}\\
    =& \phi_e^{r}(E_{i})\big(\sum\limits_{k}P^{\odot}_{ek}P^{}_{kk}P^{\oplus}_{k\beta}\big)\delta_{rs} + \int dE_{i}\phi^{r}_{e}(E_{i})P^{\odot}_{ei}P^{}_{ij}w^{}_{ij}P^{\oplus}_{j\beta},
\end{align}
with integral limits $[E_{j},E_{j}/\delta^{2}]$.

\section{Antineutrino flux}

\begin{figure}[bt]
    \centering
    \includegraphics[trim=.5cm .0cm .5cm .0cm, clip=true, width=1\linewidth]{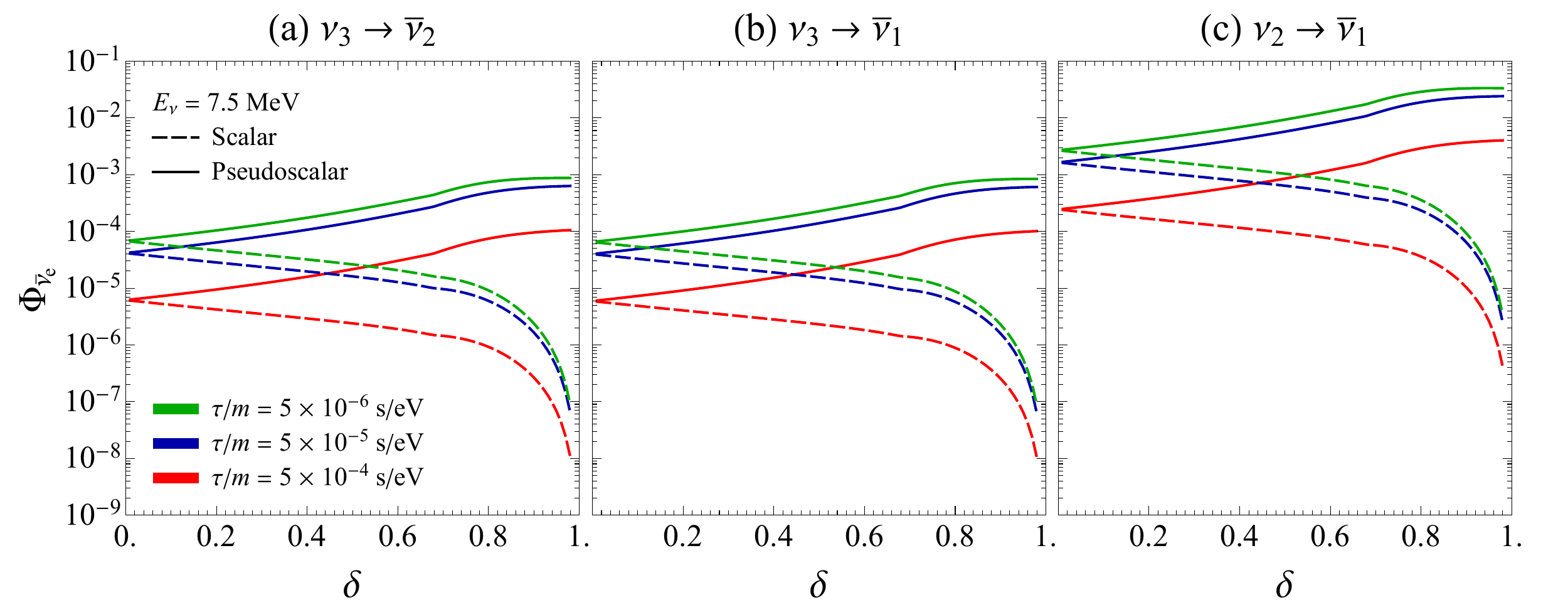}
    \caption{Solar electron antineutrino flux for different decay channels in the normal hierarchy as a function of $\delta$ for a benchmark energy of $\SI{7.5}{\mega\electronvolt}$. The red, blue and the green curves are for $\tau/m~=~\SI{5e-4}{\second\per\electronvolt}$, $\tau/m~=~\SI{5e-5}{\second\per\electronvolt}$ and $\tau/m~=~\SI{5e-6}{\second\per\electronvolt}$ respectively. The dashed and solid curves are for scalar and pseudoscalar interactions. The left, middle and the right panels are for $\nu_{3}\rightarrow\bar{\nu}_{2}$, $\nu_{3}\rightarrow\bar{\nu}_{1}$ and $\nu_{2}\rightarrow\bar{\nu}_{1}$ respectively.}
   \label{fig:flux_nh}
\end{figure}

\begin{figure}[bt]
    \centering
    \includegraphics[trim=.5cm .0cm .5cm .0cm, clip=true, width=1\linewidth]{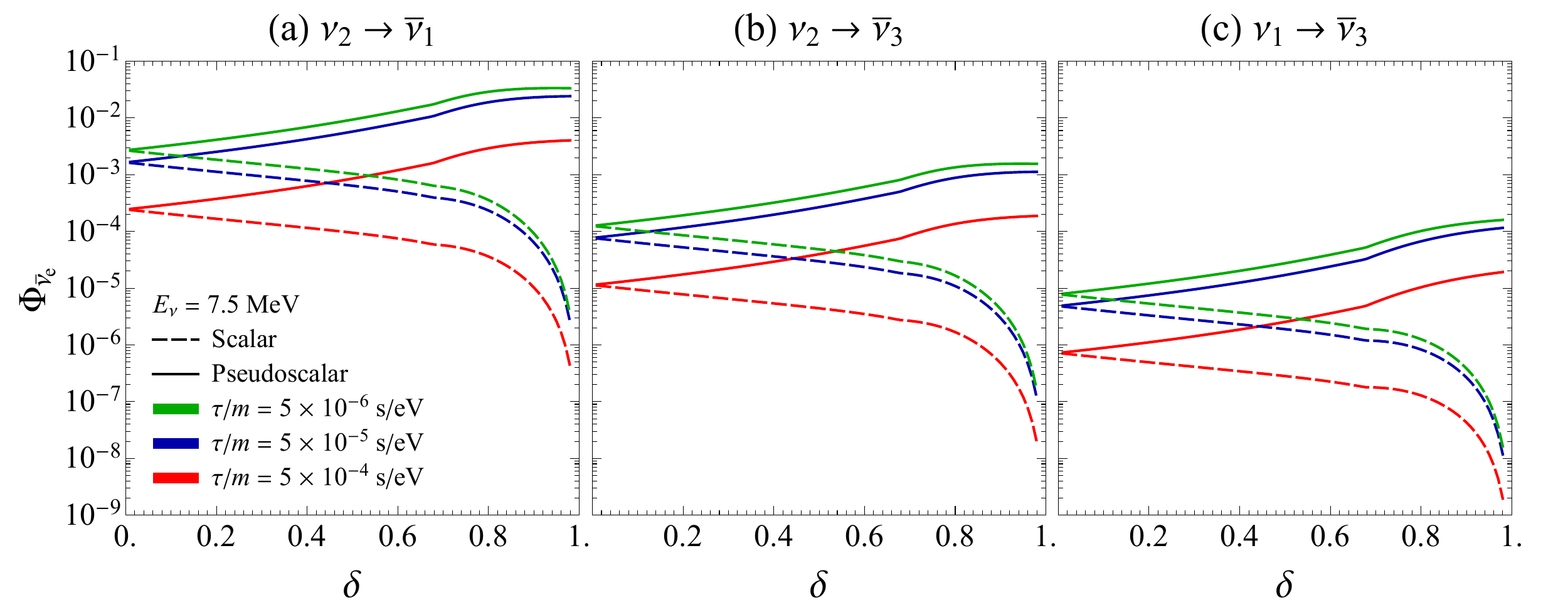}
    \caption{The same as Suppl. Fig.~\ref{fig:flux_nh} for inverted hierarchy.  The left, middle and the right panels are for $\nu_{2}\rightarrow\bar{\nu}_{1}$, $\nu_{2}\rightarrow\bar{\nu}_{3}$ and $\nu_{1}\rightarrow\bar{\nu}_{3}$ respectively. }
    \label{fig:flux_ih}
\end{figure}

In this section we present the expected $\bar{\nu}_{e}$ flux from the Sun under all possible decay channels as shown in Suppl.Fig.~\ref{fig:flux_nh} for the decays in the normal hierarchy --- $\nu_{3}\rightarrow\bar{\nu}_{2}$, $\nu_{3}\rightarrow\bar{\nu}_{1}$ and $\nu_{2}\rightarrow\bar{\nu}_{1}$ --- and Suppl.Fig.~\ref{fig:flux_ih} for the decays in the inverted hierarchy --- $\nu_{2}\rightarrow\bar{\nu}_{1}$, $\nu_{2}\rightarrow\bar{\nu}_{3}$ and $\nu_{1}\rightarrow\bar{\nu}_{3}$. Here we have chosen a benchmark energy of $\SI{7.5}{\mega\electronvolt}$ for demonstration purposes. We show both scalar and pseudoscalar interactions by the dashed and solid curves respectively. We choose three benchmark neutrino lifetime values to show the effects of the lifetime on the flux. The red, blue and the green curves are for $\tau/m~=~\SI{5e-4}{\second\per\electronvolt}$, $\tau/m~=~\SI{5e-5}{\second\per\electronvolt}$ and $\tau/m~=~\SI{5e-6}{\second\per\electronvolt}$ respectively. 

We note that the expected flux increases with decreasing decay lifetime. As we get to faster decays, we see that the change in flux is reduced because, when the decay is sufficiently fast, all neutrinos decay to antineutrinos and no more antineutrinos can be produced. We notice the probability is independent of the nature of the interactions, as can be seen from Eq.~\eqref{eq:A}. We see that the behavior is different for the scalar (solid) and pseudo-scalar (dashed) cases for $\delta\neq0$. For scalar case, the probability decreases with increasing $\delta$ and at $\delta\sim1$, the probability is zero. For pseudo-scalar interactions, the probability increases with increasing $\delta$ and goes to a constant value as $\delta\rightarrow1$.
  \begin{figure}[bt]
    \centering
    \includegraphics[trim=.5cm .0cm .5cm .0cm, clip=true, width=1\linewidth]{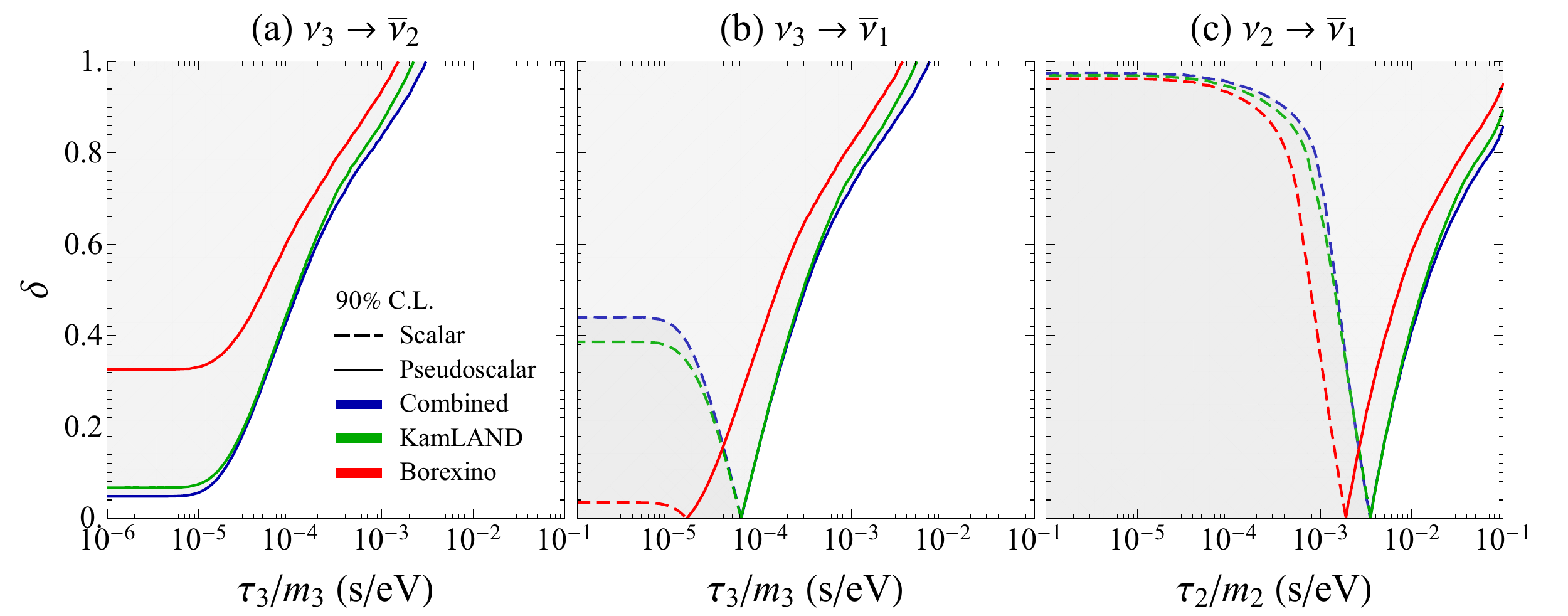}
    \caption{Limits for different experiments in the $\tau/m$ vs $\delta$ plane at $90$\% C.L. for normal hierarchy. The green, red and blue curves are for KamLAND, Borexino and their combination respectively. The shaded region is disallowed. The left, middle and the right panels are for $\nu_{3}\rightarrow\bar{\nu}_{2}$, $\nu_{3}\rightarrow\bar{\nu}_{1}$ and $\nu_{2}\rightarrow\bar{\nu}_{1}$ respectively.}
    \label{fig:chi2_nh}
\end{figure}

If we dissect various possible decay channels, we see that for $\nu_{3}\rightarrow\bar{\nu}_{1}$ and $\nu_{3}\rightarrow\bar{\nu}_{2}$ (Suppl.Fig.~\ref{fig:flux_nh}), we see that the latter case gives slightly more antineutrinos. The reason is that $P^{\oplus}_{2e}$ is bigger than $P^{\oplus}_{1e}$ for $\theta_{12}\simeq33^{\circ}$. If we compare the two cases where the daughter is $\bar{\nu}_{3}$ (Suppl.Fig.~\ref{fig:flux_ih}), we again see that $\nu_{2}\rightarrow\bar{\nu}_{3}$ gives larger antineutrino flux than $\nu_{1}\rightarrow\bar{\nu}_{3}$ channel. This can be attributed to the standard MSW effect \cite{Mikheev:1986gs}. The amount of $\nu_{2}$ is much larger than the $\nu_{1}$ in solar-neutrinos at these energies because of the MSW resonance. Hence, we see more antineutrinos from the $\nu_{2}\rightarrow\bar{\nu}_{3}$ decay. Finally, we show the $\nu_{2}\rightarrow\bar{\nu}_{1}$ decay channel in Suppl.Fig.~\ref{fig:flux_nh}. This gives the highest antineutrino flux as there is no suppression due to the $\sin^{2}\theta_{13}$ in the probability and this channel is widely studied for the solar neutrinos.  

\begin{figure}[bt]
    \centering
    \includegraphics[trim=.5cm .0cm .5cm .0cm, clip=true, width=1\linewidth]{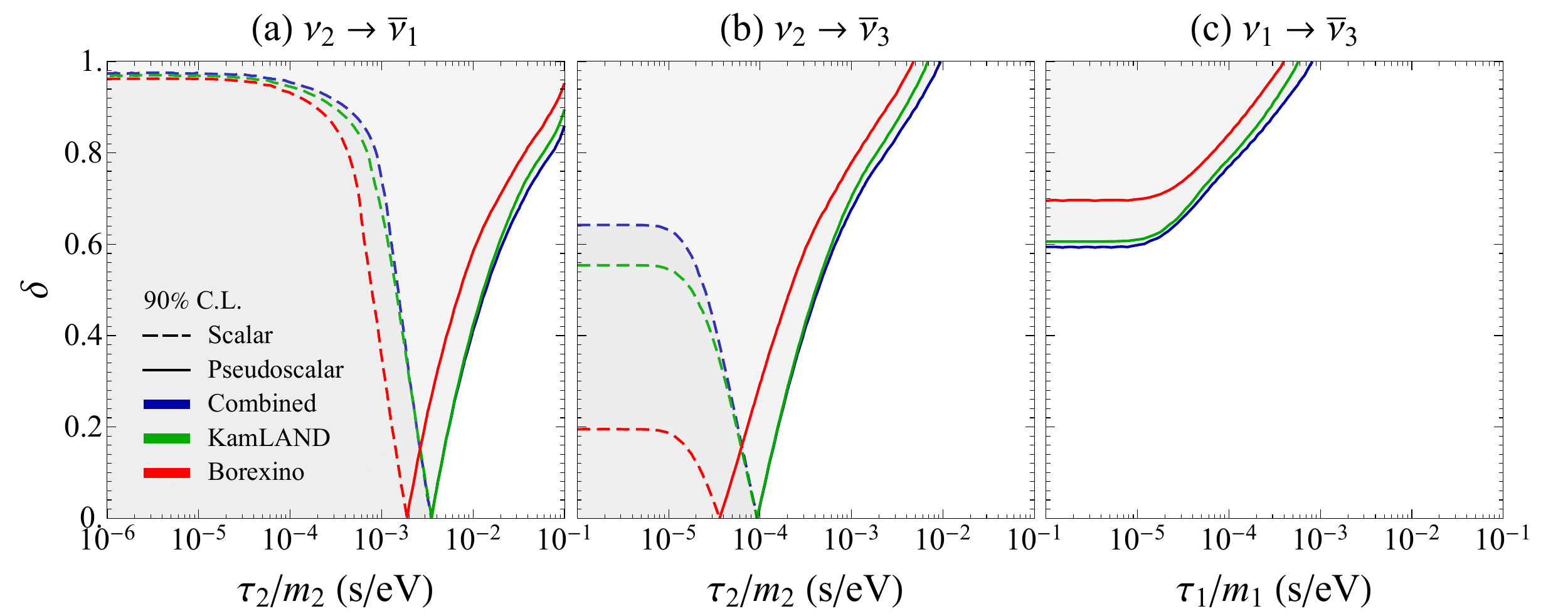}
    \caption{The same thing as Suppl.Fig~\ref{fig:chi2_nh}. for inverted hierarchy. The left, middle and the right panels are for $\nu_{2}\rightarrow\bar{\nu}_{1}$, $\nu_{2}\rightarrow\bar{\nu}_{3}$ and $\nu_{1}\rightarrow\bar{\nu}_{3}$ respectively. }
    \label{fig:chi2_ih}
\end{figure}

\section{Statistical Results}

We show limits on the decay from individual experiments separately and also their combinations for all possible neutrino-antineutrino decay channels in Suppl.Fig.~\ref{fig:chi2_nh} 
and Suppl.Fig.~\ref{fig:chi2_ih}. If neutrino has normal mass hierarchy, then $m_{3}>m_{2}>m_{1}$. In this case, three decay channels are possible, i.e., $\nu_{3}\rightarrow\bar{\nu}_{2}$, $\nu_{3}\rightarrow\bar{\nu}_{2}$ and $\nu_{2}\rightarrow\bar{\nu}_{1}$. We show these three channels in Suppl.Fig.~\ref{fig:chi2_nh}. Similarly for inverted mass hierarchy scenario, we have $m_{2}>m_{1}>m_{3}$. Again in this case, we can have three decay channels. These are $\nu_{2}\rightarrow\bar{\nu}_{1}$, $\nu_{2}\rightarrow\bar{\nu}_{3}$ and $\nu_{1}\rightarrow\bar{\nu}_{3}$. We show limits for these channels in Suppl.Fig.~\ref{fig:chi2_ih}.

We find that the data from KamLAND gives the best bounds and Borexino gives looser bounds. We also note that when KamLAND and Borexino data is added for a combined analysis, the result is marginally improved upon the KamLAND result. We find that $\nu_{3}\rightarrow\bar{\nu}_{1}$ gives strong bounds as expected and the weakest bound is obtained for the $\nu_{1}\rightarrow\bar{\nu}_{3}$. For this channel not only there is no limits for the scalar case, but also for the pseudo-scalar case, there is no limit for large fraction of the $\delta$. These results resonate with our discussions in the last sections. As we discussed in the last section, due to MSW effect, there are very few neutrinos to begin with in $\nu_{1}$ mass eigenstate.

Also, the probability is suppressed due to the small $\sin^{2}\theta_{13}$ for the $\nu_{1}\rightarrow\bar{\nu}_{3}$. Thus we get a weak bound for this case. On the other hand, due to MSW effect most of the neutrinos are in $\nu_{2}$ when they exit the Sun, therefore the channels which have parent $\nu_{2}$ provide stronger limits. And $\nu_{2}\rightarrow\bar{\nu}_{1}$ gives the strongest limits because there is no $\sin^{2}\theta_{13}$ suppression for this channel. We already discussed $\nu_{3}\rightarrow\bar{\nu}_{1}$ and $\nu_{3}\rightarrow\bar{\nu}_{2}$ channels in the main text. The $\nu_{2}\rightarrow\bar{\nu}_{3}$ gives better limits than the $\nu_{1}\rightarrow\bar{\nu}_{3}$ because again due to the MSW effect, most of the neutrinos are in $\nu_{2}$ to start with.

\begin{table}[b] 
\centering
\renewcommand{\arraystretch}{1.6}
\begin{tabular}{ l c c l }
\hline \hline
Analysis & Daughter $\nu$ included & Lower Limit  (s/eV) \\
\hline
Atmospheric and long-baseline data~\cite{GonzalezGarcia:2008ru}   & No &  $ 290 \times 10^{-12}$ (90\% C.L)\\
MINOS and T2K data~\cite{Gomes:2014yua} &   No & $ 2.8 \times 10^{-12}$ (90\%~C.L.) \\
MINOS and T2K data~\cite{Gago:2017zzy} &   Yes  &  $15 \times 10^{-12}$ (90\%~C.L.) \\
NOVA and T2K data~\cite{Choubey:2018cfz} &   No  &  $1.5 \times 10^{-12}$ (90\%~C.L.) \\
KamLand data~\cite{Porto-Silva:2020gma} & Yes & $1100\times 10^{-12}$ (90\% C.L)\\
DUNE expected sensitivity~\cite{Coloma:2017zpg} &   Yes &  $ \left(195 - 260\right) \times 10^{-12}$ (90\% C.L.) \\
DUNE expected sensitivity~\cite{Ghoshal:2020hyo} & No & $51\times 10^{-12}$ (90\% C.L) \\
ICAL expected sensitivity~\cite{SMohan:2018tch} & No & $ 160 \times 10^{-12}$ (90\% C.L)\\
T2HKK expected sensitivity~\cite{Chakraborty:2020cfu} & No & $43.6\times 10^{-12}$ ($3\sigma$ C.L)\\
ESSnuSB expected sensitivity~\cite{Choubey:2020dhw} & No & $26.4\times 10^{-12}$ ($3\sigma$ C.L)\\
MOMENT expected sensitivity~\cite{Tang:2018rer} & No & $10\times 10^{-12}$ ($3\sigma$ C.L) \\
ORCA expected sensitivity~\cite{deSalas:2018kri} & No & $140\times 10^{-12}$ (90\% C.L) \\
JUNO expected sensitivity~\cite{Abrahao:2015rba} & No & $75 \times 10^{-12}$ (95\% C.L)\\
JUNO expected sensitivity~\cite{Porto-Silva:2020gma} & Yes & $100\times 10^{-12}$ (90\% C.L)\\
This work (KamLand/Borexino/Super-Kamiokande data) & Yes &  $3\times 10^{-5}$ (90\% C.L) \\
\hline
\hline
\end{tabular}
\caption{\label{tab:bounds}
\footnotesize{Current and prospective constraints (expected sensitivities)
on neutrino lifetime from
neutrino oscillation experiments.  The lowest (highest)
value for DUNE sensitivity is for the highest (lowest)  $m_1$ lightest
neutrino mass. Inspired in the information from Ref.~\cite{Porto-Silva:2020gma}.}}
\end{table}

\section{Comparison with other bounds}

In the literature there are other bounds on the lifetime of heavier state $\nu_3$, that it can be categorized as 

\begin{enumerate}
\item bounds from atmospheric and long-baseline data~\cite{Barger:1998xk,GonzalezGarcia:2008ru,Gomes:2014yua,Gago:2017zzy} and sensitivity tests~\cite{Coloma:2017zpg,SMohan:2018tch,Tang:2018rer,deSalas:2018kri,Ghoshal:2020hyo,Mohan:2020tbi,Choubey:2020dhw,Chakraborty:2020cfu}, 
\item bounds from reactor experiments\cite{Porto-Silva:2020gma} and sensitivity tests~\cite{Abrahao:2015rba,Porto-Silva:2020gma},
\item bounds from astrophysical (solar, supernova and diffuse flux)~\cite{Bustamante:2016ciw,Delgado:2021vha} 
\end{enumerate}
also these bounds include or not include the daughter neutrinos. When did not include the daughters neutrinos the bounds applies equally to Dirac or Majorana neutrinos. When it includes the daughters neutrinos, each analysis assumed a different approach to include or not include the Majorana decay channels.

Other scenarios assume a heavy $\nu_4$ state, mostly sterile,  is decaying~\cite{Hostert:2020oui,deGouvea:2019qre} was used in respectively solar neutrinos and short-baseline neutrino experiments.

\end{document}